\documentclass[conference]{IEEEtran}
\IEEEoverridecommandlockouts

\usepackage{cite}
\usepackage{amsmath,amssymb}
\usepackage{graphicx}
\usepackage{booktabs}
\usepackage{siunitx}
\usepackage{url}
\usepackage{float}
\usepackage[colorinlistoftodos]{todonotes}

\graphicspath{{figures/}}

\newcommand{\FaultyDev}{\texttt{mblr-2/00082752}}

\newcommand{\NReadings}{\num{902396}}       
\newcommand{\NDevices}{45}                  
\newcommand{\NDevicesUsable}{41}            
\newcommand{\SizeMB}{67}                    

\newcommand{\NBuildingsMapped}{24}          
\newcommand{\NBuildings}{24}                
\newcommand{\NBuildingsOneMeter}{8}         
\newcommand{\NBuildingsTwoMeter}{15}        
\newcommand{\NBReadings}{\num{902396}}      
\newcommand{\BCompletenessOverall}{60.6}
\newcommand{\BCompletenessMedian}{65.8}

\newcommand{\BMedianSpanDays}{\num{2720}}

\newcommand{\UsableMedian}{36.1}

\newcommand{\UsableGapPts}{24}             

\newcommand{\PartialSumPct}{29.8}           
\newcommand{\NBNegDelta}{\num{113870}}      
\newcommand{\NBNegInflation}{\num{1251}}    
\newcommand{\NBNegMedianTwo}{\num{7319}}    
\newcommand{\NBNegMedianOne}{31}            

\newcommand{\DateStart}{15 January 2019}
\newcommand{\DateEnd}{4 August 2026}

\newcommand{\SpanMonths}{91}
\newcommand{\MedianSpanDays}{\num{2719}}    

\newcommand{\NExpected}{\num{2525344}}
\newcommand{\NMissing}{\num{1109219}}
\newcommand{\CompletenessOverall}{56.1}
\newcommand{\CompletenessMedian}{58.2}      
\newcommand{\CompletenessMin}{20.0}         
\newcommand{\CompletenessMax}{83.3}         
\newcommand{\CompletenessWorstAll}{4.5}     

\newcommand{\CompYearBest}{84}            
\newcommand{\CompYearBestY}{2020}
\newcommand{\CompYearWorst}{40}           
\newcommand{\CompYearWorstY}{2024}

\newcommand{\NGapsShort}{\num{212002}}      
\newcommand{\NGapsLong}{\num{3529}}         

\newcommand{\NOutages}{4}                   
\newcommand{\OutageDaysTotal}{595}          
\newcommand{\OutageMaxDays}{135}            
\newcommand{\OutageMaxHours}{\num{3242}}
\newcommand{\OutageMaxWindow}{November 2024--March 2025}
\newcommand{\OutageSecondDays}{94}          
\newcommand{\OutageThirdDays}{69}           
\newcommand{\OutageFourthDays}{61}          

\newcommand{\NNegDeltaOthers}{91}           
\newcommand{\NDevFlat}{3}                   
\newcommand{\NDevMostlyFlat}{10}            
\newcommand{\ZeroDeltaMax}{100\%}
\newcommand{\MaxVolume}{\num{645465551}}    
\newcommand{\MinReadings}{1}

\newcommand{\NDupRows}{\num{1905}}          
\newcommand{\NDupConflicting}{0}            
\newcommand{\NDevDup}{42}                   

\newcommand{\NDevStale}{16}                 

\newcommand{\FixDecreasesBefore}{\num{205200}}   
\newcommand{\FixDecreasesAfter}{0}
\newcommand{\FixDuplicateRows}{\num{1759}}       
\newcommand{\FixBuildingsBefore}{23}             
\newcommand{\FixBuildingsAfter}{24}
\newcommand{\FixVolumeRatio}{0.998}              
\newcommand{\MeterVolumeTruth}{\num{727867805}}

\newcommand{\RedistMaxDays}{7}
\newcommand{\RedistMaxRate}{20}
\newcommand{\GapBrackets}{\num{898914}}
\newcommand{\GapRejectedLong}{374}
\newcommand{\GapRejectedRate}{\num{3084}}
\newcommand{\RedistMedianGap}{4}                 
\newcommand{\RedistMaxGap}{168}                  

\newcommand{\GoldRows}{\num{1518462}}
\newcommand{\PctObserved}{53.6}
\newcommand{\PctImputed}{22.3}
\newcommand{\PctMeasured}{75.9}

\newcommand{\PctUnrecoverable}{24.1}
\newcommand{\RawAvailability}{59.4}

\newcommand{\ClusterK}{2}
\newcommand{\ClusterSil}{0.389}
\newcommand{\ClusterBig}{20}
\newcommand{\ClusterSmall}{4}
\newcommand{\ProfileMinSupport}{188}   

\newcommand{\DaiadUsers}{\num{1099}}
\newcommand{\DaiadUsersStated}{\num{1007}}

\newcommand{\DaiadCompleteness}{80.2}

\newcommand{\DaiadUsersBelowHalf}{144}

\newcommand{\BdgMeters}{\num{3053}}
\newcommand{\BdgBuildings}{\num{1636}}

\newcommand{\VolUnit}{litres}
\newcommand{\MeterResolutions}{1, 10 and 100\,\unit{\litre}}

\newcommand{\DeliveryMin}{42}      
\newcommand{\DeliveryMax}{70}

\begin{document}

\newcommand\copyrighttext{%
  \footnotesize \textcopyright 2026 IEEE. Personal use of this material is permitted. Permission from IEEE must be obtained for all other uses, in any current or future media, including reprinting/republishing this material for advertising or promotional purposes, creating new collective works, for resale or redistribution to servers or lists, or reuse of any copyrighted component of this work in other works. 
  \linebreak Preprint submitted to the Workshop on Water Supply Systems of the Future, 12th IEEE International Smart Cities Conference 2026 (ISC2 2026).
  }
\newcommand\copyrightnotice{%
\begin{tikzpicture}[remember picture,overlay]
\node[anchor=south,yshift=10pt] at (current page.south) {\fbox{\parbox{\dimexpr\textwidth-\fboxsep-\fboxrule\relax}{\copyrighttext}}};
\end{tikzpicture}%
}

\title{The Tethys Dataset: Seven Years of Hourly Smart Water Metering and a Pipeline for Making It Usable
}

\author{
\IEEEauthorblockN{Dimitrios Amaxilatis}
\IEEEauthorblockA{
\textit{Spark Works Ltd.}\\
Galway, Ireland \\
d.amaxilatis@sparkworks.net\\
ORCiD:0000-0001-9938-6211}
\and
\IEEEauthorblockN{Themistoklis Sarantakos}
\IEEEauthorblockA{
\textit{Spark Works Ltd.}\\
Galway, Ireland \\
tsarantakos@sparkworks.net\\
ORCiD:0000-0002-7517-6997}
\and
\IEEEauthorblockN{Georgios Mylonas}
\IEEEauthorblockA{
\textit{Industrial Systems Institute,}\\
\textit{Athena Research Center}\\
Patras, Greece \\
ORCiD:0000-0003-2128-720X}
\and
\IEEEauthorblockN{Ioannis Chatzigiannakis\thanks{*Corresponding author: I. Chatzigiannakis, ichatz@diag.uniroma1.it}\thanks{This work was supported by the WATERDEAL project, funded by the European Union’s Horizon Europe Research and Innovation programme under the Marie Sklodowska-Curie grant agreement No. 101299840.}}
\IEEEauthorblockA{
\textit{Sapienza University of Rome}\\
Rome, Italy \\
ichatz@diag.uniroma1.it\\
ORCiD:0000-0001-8955-9270}
}

\IEEEoverridecommandlockouts
\IEEEpubid{\makebox[\columnwidth]{
\hfill} \hspace{\columnsep}\makebox[\columnwidth]{ }}

\maketitle

\copyrightnotice

\begin{abstract}
Methods for water demand forecasting and leak detection are based on public datasets, and for water those are scarce, short, or released only after an undocumented cleaning process, hiding defects of the deployment they came from. We present \emph{Tethys}: \SpanMonths{} months of hourly water consumption data from \NBuildings{} buildings of a municipal water network, published with a quantitative account of its quality, rather than in place of one. Raw availability is \RawAvailability{}\%, while loss is not random: 4 fleet-wide outages totalling \OutageDaysTotal{} days interrupt the entire estate at once. We show that the aggregation producing the released files silently introduced \FixDecreasesBefore{} impossible decreases in a cumulative index, and that correcting it is a one-line change. Because the meters are cumulative, the readings bracketing a short gap fix the volume that passed through it, so \PctMeasured{}\% of hours rest on a measurement, while \PctUnrecoverable{}\% are reported as unknown. We release the dataset, its per-hour provenance, and the pipeline that produces it.
\end{abstract}


\section{Introduction}
\label{sec:intro}
Water utilities across Europe are 
in the process of introducing sensors to monitor their networks, and the resulting consumption data underpins demand forecasting, leak localisation and conservation policy~\cite{taloma2025machine}. Methods to tackle such tasks are being developed and compared based on public datasets, thus making such datasets critical to this research field. However, while electricity metering has long-running public corpora, such as the Building Data Genome Project 2
~\cite{bdg2}, the water equivalents are fewer, shorter, or narrower. The DAIAD release covers \num{1007} consumers, but ends after roughly two and a half years~\cite{daiad2017}, while end-use corpora such as WEUSEDTO offer high resolution over a handful of instrumented households~\cite{weusedto2022}. Multi-year records specifically from an operating municipal deployment still remain scarce.

In addition, while datasets are typically released after undergoing cleaning, the cleaning procedure itself is rarely described in detail, so the defects of the original deployment disappear from the record along with any evidence of how they were handled. Those defects are not incidental: imputation methods are usually evaluated by masking values at random in an otherwise complete matrix~\cite{saits2023}. If real loss is concentrated in correlated, system-wide outages, such an evaluation measures something the deployment will never present.

In this work, we describe in detail the \emph{Tethys} dataset: \NReadings{} hourly readings from \NBuildings{} buildings of a municipal water network, served by \NDevices{} metering endpoints, from \DateStart{} to \DateEnd{}, i.e., \SpanMonths{} months of continuous operation. Our contributions include:

\begin{itemize}
\item a description of the dataset and its acquisition, including the schema property that governs everything downstream: readings are cumulative meter indices, not per-interval consumption (Section~\ref{sec:description});
\item a quantitative data-quality characterisation covering completeness, a gap taxonomy separating benign dropouts from \NOutages{} fleet-wide outages totalling \OutageDaysTotal{} days, counter anomalies, and an aggregation defect inflating apparent resets a thousandfold (Section~\ref{sec:quality});
\item an Apache Airflow curation pipeline organised as medallion tiers, which converts the raw export into analysis-ready per-building series and refuses to impute what cannot honestly be recovered (Section~\ref{sec:pipeline});
\item an explicit description of which research tasks the dataset supports and which it cannot (Section~\ref{sec:conclusion}).
\end{itemize}

Raw data availability across the dataset is \RawAvailability{}\%, and \PctObserved{}\% of hours yield a directly observed consumption interval.
As the values reported by the meters are cumulative, readings bracketing a short gap still provide the volume that passed through it, increasing the share of measurements with at least partial data to \PctMeasured{}\%. 
The remaining \PctUnrecoverable{}\% are reported as unknown rather than filled: nothing in the released dataset is invented. 
What a deployment records and what it can support are different questions, and this paper reports both.

\section{Related Work and Comparable Datasets}
\label{sec:related}
Public water consumption datasets fall into two groups, and Tethys 
has elements of both. \emph{End-use corpora} instrument individual fixtures at very high resolution to support disaggregation research. E.g., WEUSEDTO provides 1-second resolution at fixture level, but a single apartment occupied by one person over roughly a year~\cite{weusedto2022}. While such datasets can help answer questions about appliance behaviour, they are not particularly useful in estate-scale scenarios.
\emph{Utility-scale meter releases} sit at the opposite extreme. DAIAD
published hourly readings for a large consumer sample from the AMAEM utility in Alicante~\cite{daiad2017}. It bears similarities to Tethys in terms of resolution and metering technology; however, it covers roughly a third of its time duration. 
For contrast, electricity metering has no equivalent shortage: the Building Data Genome Project 2 alone releases \BdgMeters{} meters across \BdgBuildings{} buildings~\cite{bdg2}. 

This asymmetry is what motivates us to release a water dataset of comparable longitudinal depth.
At \SpanMonths{}~months long, Tethys is roughly three times the span of DAIAD and more than three times than BDG2. Additionally, the two largest fleet-wide outages both occur after 2023, so a record of DAIAD's length would not have exposed them. Completeness, meanwhile, is almost never reported. None of the compared datasets publishes it, so we measured it for DAIAD from the released file, obtaining \DaiadCompleteness{}\% overall against each user's own hourly grid and \DaiadUsersBelowHalf{} users below 50\%. DAIAD is cleaner than Tethys, which we state plainly since the file is public, but the figure is one a user must derive themselves, and doing so surfaced a discrepancy: the release describes \DaiadUsersStated{} consumers while the file contains \DaiadUsers{} distinct user keys. Quantities that go unreported also go unchecked.

\section{The Tethys Deployment}
\label{sec:deployment}
The installation's design and instrumentation are described in
detail in \cite{app10061965}; we summarise here only what bears on the
measurements. Consumption meters of \MeterResolutions{} resolution broadcast over \textit{wM-Bus} to a gateway in the building they serve, which forwards data over a private LoRaWAN network at 868MHz to edge nodes that communicate with the platform's cloud services. The recorded index is therefore in \VolUnit{}, at a granularity that differs between meters, and the export analysed here is drawn
from the cloud layer.

Two features of the installation explain most of what
Section~\ref{sec:quality} measures. The meters sit in cast iron-covered
inspection pits below their buildings, and gateways were placed on upper floors to compensate for signal attenuation. Even so, per-device delivery of expected hourly broadcasts was measured at \DeliveryMin{}--\DeliveryMax{}\%, degraded by the pit enclosures, antenna displacement from wind and birds, and background noise from unrelated 868MHz equipment~\cite{app10061965}. Our completeness figures, taken seven years later from the stored data, agree
closely with those radio-layer rates, so the gaps are predominantly transmission losses rather than metering failures. That distinction matters: a lost packet leaves the meter's own counter intact, which is what Section~\ref{sec:gap recovery} exploits.
Monitored buildings are non-residential premises served by either one or two meters, with the total building's consumption counted as the sum of the installed meters; no personal data is collected as the consumption refers to the building's aggregated hourly water intake.

\section{Dataset Description}
\label{sec:description}

\subsection{Schema and Format}
Tethys is distributed as one CSV file per building on an hourly grid, using the fields of Table~\ref{tab:schema}. Consumption is the volume for each hour in \VolUnit{}, while two fields state where it came from: \texttt{method} distinguishes a directly observed hour from one holding a share of a total measured across a gap, and \texttt{gap\_hours} gives the length of that gap. An hour whose consumption is unknown is null and labelled accordingly.

\begin{table}[t]
\caption{Released record schema, with a representative row.}
\label{tab:schema}
\centering
\footnotesize
\begin{tabular}{@{}lll@{}}
\toprule
\textbf{Field} & \textbf{Type} & \textbf{Example} \\
\midrule
timestamp   & int64 (epoch ms, UTC) & \texttt{1550826000000} \\
datetime    & ISO 8601, UTC         & \texttt{2019-02-22T09:00:00Z} \\
consumption & float (\VolUnit{})    & \texttt{560.0} \\
method      & enum                  & \texttt{imputed} \\
gap\_hours  & int                   & \texttt{6} \\
\bottomrule
\end{tabular}
\end{table}

The underlying meters report a \emph{cumulative} index rather than per-interval consumption, and that property governs everything downstream. Consumption is recovered by first-differencing, which is not a neutral operation: it turns every missing reading into a corrupted interval rather than an absent one, and any counter reset into a large negative value. Section~\ref{sec:pipeline} describes how the released files are derived, and we publish the differenced series precisely so that consumers need not repeat that reasoning. The nominal sampling interval is one reading per hour, and meter resolution varies across the fleet, so the smallest resolvable quantity differs between buildings.

\subsection{Volume and Coverage}
The released dataset comprises \NReadings{} building-hours across
\NBuildings{} buildings, spanning \DateStart{} to \DateEnd{} (\SpanMonths{} months) and occupying approximately \SizeMB{}\,MB as raw CSV. Coverage is long rather than wide: the median endpoint record spans \MedianSpanDays{} days, so the dataset is dominated by a core of endpoints observed across the full period. More than 7 years of continuous operation is its main distinguishing feature against the datasets surveyed in Section~\ref{sec:related}, which are typically an order of magnitude shorter.

\subsection{From Endpoints to Buildings}
Endpoints are not the natural unit of analysis, since consumption attributable to a building is the sum over the meters serving it. Of the \NBuildingsMapped{} buildings, \NBuildings{} are analysed, \NBuildingsOneMeter{} served by a single meter and \NBuildingsTwoMeter{} by two; one is excluded, served solely by \FaultyDev{}, a persistently malfunctioning endpoint discussed in Section~\ref{sec:quality}.
A one-character error in the released map excludes a further building, which we correct.

\begin{figure}[t]
\centering
\includegraphics[width=\linewidth]{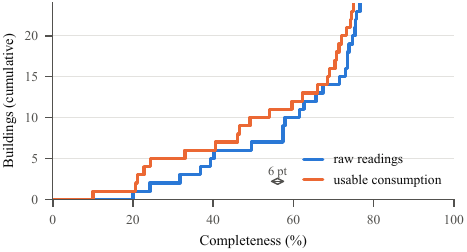}
\caption{Per-building completeness: raw readings versus usable consumption. The \UsableGapPts{}-point gap is the cost of requiring a valid interval rather than a present reading.}
\label{fig:cdf}
\end{figure}

At building level, the dataset holds \NBReadings{} building-hours with at least one meter reporting, giving \BCompletenessOverall{}\% raw availability (median \BCompletenessMedian{}\%) over a median span of \BMedianSpanDays{}~days. Raw availability counts hours in which a reading exists; \emph{usable} coverage counts hours yielding a valid consumption interval, requiring 2 consecutive readings an hour apart, no reset between them, and every meter present. By this measure, the median building falls to \UsableMedian{}\%, a gap of roughly \UsableGapPts{}~points shown in Fig.~\ref{fig:cdf} and accounted for in Section~\ref{sec:quality}. The two are easily conflated, and only the second bounds what the data can support.

\begin{figure*}[t]
\centering
\includegraphics[width=0.93\textwidth]{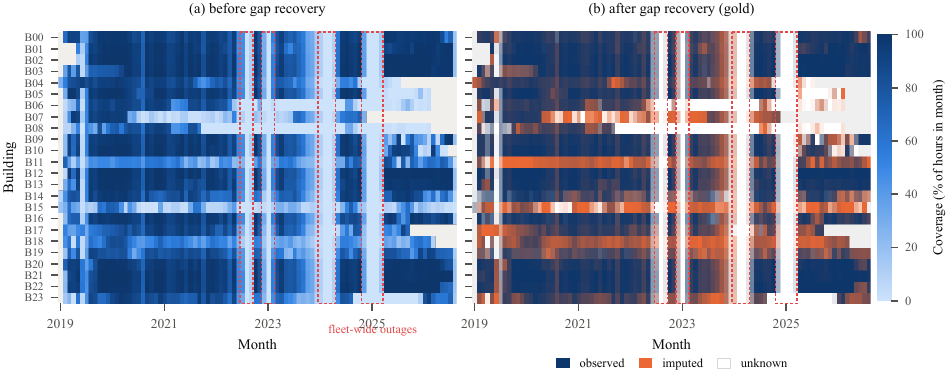}
\caption{Monthly coverage per building, before (a) and after (b) curation, on
the same criterion. Cells in (b) are shaded by provenance: \PctObserved{}\% of
hours are direct readings and \PctImputed{}\% a share of a measured total
spread over at most \RedistMaxDays{}~days. Nothing is interpolated, so pale
cells are unknown rather than filled, and the dashed fleet-wide outages close in
volume but not in hourly detail. Grey cells lie outside a building's service
period.}
\label{fig:heatmap}
\end{figure*}

\subsection{Consumption Characteristics}
Differenced consumption shows the pattern expected of institutional premises,
rising sharply from around 05:00, sustaining through the working day and
collapsing at weekends, though some buildings instead hold a near-constant level
with an alternating hour-to-hour pattern that reflects a two-hourly reporting
regime rather than demand. Absolute magnitudes differ by several orders of
magnitude across the estate, so aggregate error metrics computed over all
buildings are dominated by the largest consumers and per-building normalisation
is needed before any pooled model is evaluated.

\section{Data Quality Characterisation}
\label{sec:quality}

All figures in this section are computed directly from the raw export, by
comparing observed readings against the hourly grid implied by each endpoint's
own first and last reading. Measuring against a global calendar would penalise
endpoints commissioned late or decommissioned early for absences that are not
data-quality failures. We first
remove \NDupRows{} exact duplicate records, affecting \NDevDup{} endpoints; all
duplicate groups carry identical values (\NDupConflicting{} conflicting cases),
so removal is lossless, but retaining them would inflate completeness.

\subsection{Completeness}
Across the fleet, \CompletenessOverall{}\% of expected hourly slots are
present: \NMissing{} of \NExpected{} readings are absent. Completeness is also unevenly distributed. Among endpoints with a substantive record
(\NDevicesUsable{} of \NDevices{}), the median is \CompletenessMedian{}\% and values range from \CompletenessMin{}\% to \CompletenessMax{}\%; the weakest endpoint overall retains only \CompletenessWorstAll{}\% of its expected readings, and the least productive contributes just \MinReadings{} readings in total.

The distribution matters more than the headline; even the best-performing endpoint attains \CompletenessMax{}\% completeness, i.e., there is no clean subset of the fleet: 
an analysis restricted to well-behaved endpoints would discard the entire dataset. Missingness must be treated as a modelling problem, not as a preprocessing nuisance. Coverage also degrades with deployment age, 
peaking at \CompYearBest{}\% in
\CompYearBestY{} and falling to \CompYearWorst{}\% by \CompYearWorstY{} before partly recovering, while \NDevStale{} of \NDevices{} endpoints fall silent more than 90 days before the record ends.

\subsection{Gap Structure}
Aggregate completeness understates the problem, because the missing readings are not scattered independently. Gaps fall into 3 classes, and class G3 is identified by testing whether per-endpoint gaps \emph{overlap in calendar time} rather than by grouping endpoints whose gaps happen to be of similar duration: equal-length gaps at unrelated times are not a shared event, and the distinction
changes the conclusion. G1, short dropouts of at most 24\,h, accounts for \NGapsShort{} events and is adequately handled by local interpolation. G2, extended per-endpoint outages beyond 24\,h, accounts for \NGapsLong{} more and is recoverable from contemporaneous peers. G3, fleet-wide outages, comprises \NOutages{} events totalling \OutageDaysTotal{} days for which no donor series exists at all.

Class G3 is the substantive finding. Counting hours in which at least 90\% of
active endpoints are simultaneously silent yields \OutageDaysTotal{} such days
across the record, concentrated in \NOutages{} contiguous episodes exceeding
one month. The longest spans \OutageMaxHours{}\,hours, or \OutageMaxDays{}~days
across \OutageMaxWindow{}, followed by episodes of
\OutageSecondDays{}, \OutageThirdDays{} and \OutageFourthDays{} days. An
interruption shared across the fleet cannot originate at the meters; it
reflects failure in the shared ingestion path.

The consequence is structural. Methods that reconstruct a missing value by
borrowing from correlated series observed at the same instant have nothing to
borrow from, because every candidate donor is missing simultaneously. Random
masking of a complete matrix, the standard protocol for evaluating imputation,
cannot produce a gap of this shape, and results obtained under that protocol
therefore do not transfer to this deployment~\cite{saits2023}. We note
that the two largest episodes both fall after 2023, so an extract truncated at
that point would not reveal the failure mode at its worst.

\subsection{Counter Anomalies}
Because readings are cumulative, the integrity of the counter determines the integrity of the derived consumption series, 
with two potential issues:

\emph{Stalled counters.} \NDevFlat{} endpoints report a constant value for
their entire record, and \NDevMostlyFlat{} have zero-difference fractions above
\num{0.9}, reaching \ZeroDeltaMax{}. A stalled counter and a genuinely idle
building are numerically identical in this schema. Nothing in the data separates
them and no ground truth exists against which a separation could be validated.

\emph{Implausible magnitudes.} One endpoint's entire record consists of a single
reading of \MaxVolume{}, a value inconsistent with any physical measurement at
this scale and better explained as a sentinel or field overflow. Such values are
not rejected at the source, and because a cumulative index is summed rather than
averaged, one sentinel offsets a building's whole series by its magnitude.
Range-checking before aggregation is therefore not optional.

The export analysed here contains no negative first differences at all, so
counter resets are absent from the released record. Earlier extracts of the same
deployment did contain them, and a pipeline consuming future exports cannot
assume their absence; Section~\ref{sec:pipeline} accordingly retains
reset detection.

\subsection{Aggregation Artefacts}
\label{sec:merge}
A final defect comes not from the meters but from the device-to-building
aggregation, which summed cumulative indices per timestamp. That is valid only
when every meter reports, so whenever one of two was absent the total dropped by
its whole accumulated index, and \PartialSumPct{}\% of readings in two-meter
buildings were such partial sums. Negative first differences consequently rise
from \NNegDeltaOthers{} at meter level to \NBNegDelta{} at building level, an
inflation of roughly \NBNegInflation{}$\times$, with a median of
\NBNegMedianTwo{} for two-meter buildings against \NBNegMedianOne{} for
single-meter ones. Virtually every apparent reset in the aggregated series was therefore an artefact of the merge, and it also explains why that series looked more complete: it accepted readings that were incomplete sums. Section~\ref{sec:fixagg} gives the correction.

\section{A Curation Pipeline for Tethys}
\label{sec:pipeline}
Characterising the defects of Section~\ref{sec:quality} is only part of the contribution. Each one implies a specific remedy, and we implement those as a reproducible Apache Airflow pipeline. 
Publishing it alongside the data 
is important, since the deployment is live, and future exports require similar treatment.

\subsection{Correcting the Aggregation}
\label{sec:fixagg}
The pipeline consumes the per-building files the project already produces, assuming such files are correct, which they were not. Every defect came from one function summing cumulative indices as though they were independent. A meter that failed to report contributed nothing, so the building total fell by that meter's whole accumulated index; since an absent meter still stands at its last known value, carrying that value forward before summing removes \FixDecreasesBefore{} spurious decreases, leaving \FixDecreasesAfter{}. Three smaller faults compound it: duplicate records were added twice (\FixDuplicateRows{} rows), the first line of every file was skipped as a header though these files have none, and a one-character error in the device map excluded a building, so \FixBuildingsBefore{} were published where \FixBuildingsAfter{} exist.

Correctness here is verifiable rather than asserted. The per-meter export fixes the total volume recorded at \MeterVolumeTruth{} \VolUnit{}, and the corrected aggregation reproduces \FixVolumeRatio{} of it against a series that is non-decreasing throughout. We also emit the number of meters reporting each hour, which makes the files self-describing: a consumer can distinguish a directly measured hour from one holding a carried-forward component without returning to the per-meter data. The pipeline then promotes the result through medallion tiers, each a file on disk, so any stage can be inspected or re-run
and a disputed number traced to its origin.

\subsection{Recovering Gap Totals}
\label{sec:gap recovery}
The cumulative encoding that causes such trouble also carries a property worth exploiting. A gap destroys the \emph{distribution} of consumption across its hours but not the \emph{total}, because the readings bracketing it differ by exactly the volume that passed through the meter. We therefore spread each bracketed total across the hours it spans, in proportion to the building's own hour-of-week shape, so the total is reproduced exactly and only its shape is
inferred.

The limits on this are important: 
a bracket is refused when it spans a reset, when it implies more than \RedistMaxRate{} times the building's median hourly consumption, or when it exceeds \RedistMaxDays{}~days. The last is the substantive one. Nothing prevents spreading a 3-month total across three
months of hours, and an earlier version of this work did so, but beyond about a week the hourly placement rests entirely on the profile while still being presented as measured volume. Capping at \RedistMaxDays{}~days moves those hours out of the measured category and into the unknown one: of \GapBrackets{} brackets, \GapRejectedLong{} are refused as too long and \GapRejectedRate{} as implausibly fast, while each surviving imputed hour sits within a gap of at most \RedistMaxGap{}~hours, with a median of \RedistMedianGap{}.

\subsection{What the Pipeline Refuses to Do}
No value in the gold tier is invented. An hour is emitted only when a counter measured the volume behind it, either directly or as a bracketed total, and every other hour is reported as unrecoverable. Interpolation is available in the implementation but disabled, because the hours it would fill are precisely those whose bracket has already been refused as untrustworthy; filling them would substitute a guess for evidence just rejected.

Table~\ref{tab:provenance} displays the result. Raw availability in the export is \RawAvailability{}\%, and of the \GoldRows{} hours on the gold grid, \PctObserved{}\% are direct readings and \PctImputed{}\% carry a share of a measured total, so \PctMeasured{}\% rests on a measurement. The remaining \PctUnrecoverable{}\% is null. A pipeline willing to interpolate across a multi-month outage could have reported near-total coverage; that figure would have described the method rather than the deployment. Every record carries its method and the length of the gap it came from, so a consumer can choose a standard of evidence. A study of demand volumes may use all \PctMeasured{}\%, while a study concerned with hourly peaks should keep to the \PctObserved{}\% observed directly.

\begin{table}[t]
\caption{Provenance of every hour in the gold tier.}
\label{tab:provenance}
\centering
\footnotesize
\begin{tabular}{@{}lrl@{}}
\toprule
\textbf{Provenance} & \textbf{Share} & \textbf{Volume is} \\
\midrule
observed      & \PctObserved{}\%      & measured, hourly \\
imputed & \PctImputed{}\% & measured in total, split inferred \\
unrecoverable & \PctUnrecoverable{}\% & unknown, left null \\
\bottomrule
\end{tabular}
\end{table}

\subsection{Profiling and Clustering}
The pipeline also profiles each building and clusters the shapes, which yields a negative result worth reporting. Ten scale-independent descriptors of the mean daily profile feed a k-means fit; \ClusterK{} clusters emerge with a silhouette of \ClusterSil{}, separating \ClusterSmall{} buildings with sharp early peaks and near-zero night draw from the remaining \ClusterBig{}. The intended use was to lend a sparsely observed building the shape of its neighbours.

It never activates. Over \SpanMonths{} months the least-observed hour-of-week cell
anywhere in the fleet still rests on \ProfileMinSupport{} observations, so every
building is its own best reference and the pooled shape is never needed.
Clustering-assisted imputation is therefore a function of record length rather
than of similarity between buildings: it would matter for a newly commissioned
meter, and does not for a deployment observed this long.

\section{Availability}
\label{sec:availability}
The dataset is released on Zenodo under CC~BY~4.0, and the curation pipeline under an open source licence in the accompanying repository.\footnote{DOI and repository URL to be inserted on acceptance.} We publish building-level series rather than individual meter readings. That choice is only defensible because of the correction in
Section~\ref{sec:fixagg}: the aggregation now preserves the monotonicity of the cumulative index and records how many meters contributed to each hour, so a consumer can recover consumption and judge its provenance without the per-meter data. The files released before that correction did not support either, which is the practical argument for treating aggregation as part of the dataset itself.

Each building is a CSV on the complete hourly grid of Table~\ref{tab:schema}. The \texttt{method} field is \texttt{observed}, \texttt{imputed} where the hour holds a share of a total measured across a gap of at most \RedistMaxDays{}~days, or \texttt{unrecoverable} where the value is null, and \texttt{gap\_hours} gives the length of the gap a imputed value came from, so a consumer can set their own tolerance rather than accept ours. Buildings carry anonymised keys; the readings are aggregate non-residential consumption containing no personal data, and neither the utility nor the city is named.

The pipeline reproduces the released files from the raw export in a single command, and the figures and statistics in this paper are generated from that same run. Future exports extend the record without altering its schema.

\section{Discussion and Conclusions}
\label{sec:conclusion}
Tethys records \SpanMonths{} months of a 
urban deployment, and its value lies in being reported as it is, rather than as a curated extract. 
Its missingness is structured rather than synthetic, making it a realistic setting for imputation research: the \NOutages{} fleet-wide outages interrupt every building at once, so no contemporaneous donor exists and the situation cannot be reproduced by masking values at random. It supports multi-year forecasting under genuine operating conditions, and unsupervised anomaly detection in which the absence of labels is a premise rather than an obstacle. The per-hour provenance makes the standard of evidence a deliberate choice: studies of missingness should work from the observed hours and treat the rest as the phenomenon under study, while studies of demand volume may use every measured hour.

Four boundaries follow from how the deployment was built, rather than from any fault in it. Nothing external confirms a reconstructed value, so the stalled-counter ambiguity 
cannot be settled from within the data. Buildings carry only an anonymised key, so their heterogeneity can be measured, but not explained. Hourly resolution is well-matched to demand modelling and too coarse for appliance-level disaggregation. Due to both the fleet and the ingestion path changing across the record, studies should state the extract period they used.

Two findings generalise beyond this deployment. Aggregation belongs to the dataset and deserves the scrutiny given to the measurements: one line summing cumulative indices as though they were independent introduced \FixDecreasesBefore{} impossible decreases, more than any sensor fault produced in 7 years.
Also, a cumulative counter loses the distribution of consumption across a gap but not its total, which is worth exploiting up to the point where the distribution becomes the answer rather than the question. We draw that line at \RedistMaxDays{}~days and report \PctUnrecoverable{}\% of hours as unknown, in preference to a fuller-looking dataset that would describe our method rather than the deployment. We would encourage benchmarks built on this data to state which provenance they admit, and imputation studies to test against its fleet-wide outages rather than against randomly masked values.

\bibliographystyle{IEEEtran}
\bibliography{refs}

\end{document}